\documentclass[12pt]{article}
\usepackage{axodraw,a4wide,epsfig}
\catcode`@=11
\newcount\@tempcntc
\def\@citex[#1]#2{\if@filesw\immediate\write\@auxout{\string\citation{#2}}\fi
  \@tempcnta\z@\@tempcntb\m@ne\def\@citea{}\@cite{\@for\@citeb:=#2\do
    {\@ifundefined
       {b@\@citeb}{\@citeo\@tempcntb\m@ne\@citea\def\@citea{,}{\bf ?}\@warning
       {Citation `\@citeb' on page \thepage \space undefined}}%
    {\setbox\z@\hbox{\global\@tempcntc0\csname b@\@citeb\endcsname\relax}%
     \ifnum\@tempcntc=\z@ \@citeo\@tempcntb\m@ne
       \@citea\def\@citea{,}\hbox{\csname b@\@citeb\endcsname}%
     \else
      \advance\@tempcntb\@ne
      \ifnum\@tempcntb=\@tempcntc
      \else\advance\@tempcntb\m@ne\@citeo
      \@tempcnta\@tempcntc\@tempcntb\@tempcntc\fi\fi}}\@citeo}{#1}}
\def\@citeo{\ifnum\@tempcnta>\@tempcntb\else\@citea\def\@citea{,}%
  \ifnum\@tempcnta=\@tempcntb\the\@tempcnta\else
   {\advance\@tempcnta\@ne\ifnum\@tempcnta=\@tempcntb \else \def\@citea{--}\fi
    \advance\@tempcnta\m@ne\the\@tempcnta\@citea\the\@tempcntb}\fi\fi}
\catcode`@=12

\begin{document}
\newcommand{\be}{\begin{equation}}
\newcommand{\ee}{\end{equation}}
\newcommand{\bfm}[1]{\mbox{\boldmath$#1$}}
\newcommand{\bff}[1]{\mbox{\scriptsize\boldmath${#1}$}}

\newcommand{\al}{\alpha}
\newcommand{\bt}{\beta}
\newcommand{\lm}{\lambda}
\newcommand{\bea}{\begin{eqnarray}}
\newcommand{\eea}{\end{eqnarray}}
\newcommand{\gm}{\gamma}
\newcommand{\Gm}{\Gamma}
\newcommand{\dl}{\delta}
\newcommand{\Dl}{\Delta}
\newcommand{\ep}{\epsilon}
\newcommand{\kp}{\kappa}
\newcommand{\Lm}{\Lambda}
\newcommand{\om}{\omega}
\newcommand{\pa}{\partial}
\newcommand{\nn}{\nonumber}
\newcommand{\dd}{\mbox{d}}
\newcommand{\uk}{\underline{k}}
\newcommand{\gsim}{\;\rlap{\lower 3.5 pt \hbox{$\mathchar \sim$}} \raise 1pt \hbox {$>$}\;}
\newcommand{\lsim}{\;\rlap{\lower 3.5 pt \hbox{$\mathchar \sim$}} \raise 1pt \hbox {$<$}\;}
\newcommand{\xp}{x_+}
\newcommand{\xm}{x_-}

\title{
\begin{flushright}
{\normalsize  ALBERTA-THY-12-07\\[0mm]}
{\normalsize TTP07-22\\[10mm] }
\end{flushright}
\vspace{1.0cm}
Next-to-Next-to-Leading Electroweak Logarithms in $W$-pair Production
at ILC}

\author{
  {\large J.H.~K\"uhn} $^a$,\,
  {\large F.~Metzler} $^a$,\,
  {\large A.A.~Penin} $^{b,a,c}$\\[2mm]
  $^a${\small {\it Institut f{\"u}r Theoretische Teilchenphysik,
  Universit{\"a}t Karlsruhe}}\\
  {\small {\it 76128 Karlsruhe, Germany}}\\[2mm]
  {\small\it $^b$ Department of Physics,
  University Of Alberta}\\
  {\small \it Edmonton, AB T6G 2J1, Canada}\\[2mm]
  {\small\it $^c$ Institute for Nuclear Research,
    Russian Academy of Sciences,}\\
  {\small\it  117312 Moscow, Russia}
        }
\date{}
\maketitle
\begin{abstract}
  We derive the high energy asymptotic behavior of gauge boson
  production cross section in a spontaneously broken $SU(2)$ gauge theory
  in the next-to-next-to-leading logarithmic approximation.  On the
  basis of this result we obtain the logarithmically enhanced two-loop
  electroweak corrections to the differential cross section of $W$-pair
  production at ILC/CLIC up to the second power of the large logarithm.
  \\[2mm]
\end{abstract}


\section{Introduction}
The $W$-pair production at $e^+e^-$ colliders plays a crucial role
for testing the Standard Model of electroweak interactions.  At LEP2
this process has been used for the determination of the $W$-boson
mass $M_W$, a fundamental parameter of the standard model, through
$W$-boson reconstruction with an uncertainty of 40~MeV \cite{LEP}.
Furthermore, the triple gauge boson coupling as predicted by the
non-Abelian gauge theory has been verified within a few percent. The
experimental study of the $W$-pair production at the International
Linear Collider (ILC) is expected to improve the accuracy of the
mass determination to 7~MeV due to much higher luminosity
\cite{Agu}. Moreover, the advent of ILC will give access to the new
high energy domain where the cross section is increasingly sensitive
to the triple gauge boson coupling and  $W$-pair production could be
used as a probe of the non-Abelian structure of the electroweak
interactions and of possible  gauge boson anomalous couplings. To
match the experimental accuracy, the theoretical analysis has to
take into account the electroweak radiative corrections. The
one-loop corrections to the cross section of the on-shell $W$-pair
production have been evaluated by different groups
\cite{LemVel,Boh,FJZ,Bee} already decades ago.  The calculation of
the one-loop  corrections to the $W$-boson mediated $e^+e^-\to 4f$
processes has been performed in the double pole approximation in
Ref.~\cite{Bee2} and incorporated into the event generators YFSWW
\cite{Jad} and RacoonWW \cite{Den2}. Recently the full analysis has
been completed \cite{Den}. These results ensure an accuracy
significantly better than one percent when the characteristic energy
$\sqrt{s}$ of the process is about the gauge boson mass.  However,
once $\sqrt{s}$ is far larger than $M_W$, the cross section receives
virtual corrections enhanced by powers of electroweak ``Sudakov''
logarithm $\ln\bigl({s/
  M_{W,Z}^2}\bigr)$, which at the energies of about one TeV have to be
controlled to two loops in order to keep the theoretical error below one
percent.  This is even more valid for energies of 3 TeV anticipated
for the CLIC project \cite{Ass}.
In the case of  light fermion pair production these
corrections are already available through the
next-to-next-to-next-to-leading logarithmic  (${\rm N}^3 {\rm LL}$)
approximation, {\it i.e.}
including all the two-loop logarithmic terms
\cite{Fad,KPS,KMPS,FKPS,JKPS}.  For $e^+e^-\to W^+W^-$ production only
the leading-logarithmic (LL) and the next-to-leading logarithms (NLL)
are known so far \cite{Fad,Mel1,DMP,BRV}.

In the present paper we extend the analysis of $W$-pair production to
the next-to-next-to-leading logarithmic (NNLL) approximation following
the approach developed in Refs.~\cite{KPS,KMPS,FKPS,JKPS} for the
four-fermion processes.  The limit of the small-angle production, which
could be interesting in the case of the transverse gauge bosons because
the corresponding cross section is peaked in the forward direction,
remains beyond the scope of the present paper. In the next section we
outline the approach and derive the NNLL corrections to the differential
cross section of the gauge boson pair production in a spontaneously
broken $SU(2)$ model which emulates the massive gauge boson sector of
the Standard Model of electroweak interactions.  We generalize the
result to the $SU(2)\times U(1)$ gauge theory with a heavy top quark in Section~\ref{sec3}.
A brief summary and conclusions are given in Section~\ref{sum}.

\section{High energy asymptotic of the massive gauge boson production}
\label{sec2} Let us  consider as a toy model  the spontaneously
broken $SU(2)$ gauge theory with the Higgs mechanism for gauge boson
mass generation and with  massless left-handed fermion doublets. The
model retains the main features of the massive gauge boson sector of
the Standard Model. In this case the result can be presented in a
simple analytical form and constitutes the basis for the further
extension to the full electroweak theory. We study the process of
gauge boson pair production in  fermion-antifermion annihilation at
high energy and fixed angle when all the kinematical invariants are
of the same order and  far larger than the gauge boson mass,
$|s|\sim |t| \sim |u| \gg M^2$. In this limit the asymptotic energy
dependence of the field amplitudes is dominated by {\em Sudakov}
logarithms  \cite{Sud,Jac} and  governed by the evolution equations.
The method of the evolution equations in the context of the
electroweak corrections is described in detail for  the fermion pair
production in Ref.~\cite{JKPS}. The derivation of the evolution
equations \cite{Mue,Col,Sen1} applies to any process of wide-angle
production or scattering of on-shell particles when the
characteristic momentum scale is far larger than the mass scale. It
is entirely based on (i) the properties of hard momentum region and
(ii) ultraviolet renormalization of the light-cone Wilson loops.
Therefore it depends neither  on  specific infrared structure of the
model nor on the specific choice of the external particles (for the
extension to the processes with arbitrary number of external
particles see Ref.~\cite{DJP}). Thus, the approach of
Ref.~\cite{JKPS} directly extends to the gauge boson production as
briefly described below. The only potential subtle point in the
analysis of gauge boson production is that the effects of
spontaneous symmetry breaking can change the asymptotic states as it
happens with photon and Z-boson in the standard model. This would
require an additional consideration. We, however, restrict the
analysis to the production of $W$-bosons which have the same gauge
quantum numbers in broken and symmetric phases and do not encounter
this problem.

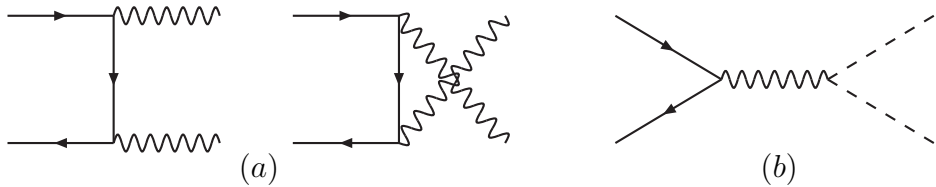
\begin{figure}[t]
\begin{center}
\hspace*{10mm}
\begin{picture}(100,60)(0,0)
\SetScale{0.8} \SetWidth{1} \ArrowLine(0,60)(50,60)
\ArrowLine(50,0)(0,0) \ArrowLine(50,60)(50,0)
\Photon(50,60)(100,60){4}{6.5} \Photon(50,00)(100,00){4}{6.5}
\Text(95,-10)[cc]{$(a)$}
\end{picture}
\hspace*{0mm}
\begin{picture}(100,60)(0,0)
\SetScale{0.8} \SetWidth{1} \ArrowLine(0,60)(50,60)
\ArrowLine(50,0)(0,0) \ArrowLine(50,60)(50,0)
\Photon(50,60)(100,00){4}{8.5} \Photon(100,60)(50,0){4}{8.5}
\end{picture}
\hspace*{5mm}
\begin{picture}(150,60)(0,0)
\SetScale{0.8} \SetWidth{1} \Photon(50,30)(100,30){4}{6.5}
\ArrowLine(50,30)(0,0) \ArrowLine(0,60)(50,30)
\DashLine(100,30)(150,0){5} \DashLine(100,30)(150,60){5}
\Text(62,-10)[cc]{$(b)$}
\end{picture}
\end{center}
\caption{\label{fig1} \small  The diagrams represent $(a)$
transverse and $(b)$ longitudinal gauge boson pair production in
fermion-antifermion annihilation at high energy in the Born
approximation.}
\end{figure}
Due to helicity conservation a pair of either transverse or longitudinal
gauge bosons can be produced in the high energy limit.
The transverse gauge bosons
behave like vector particles in the adjoint representation
while the longitudinal gauge bosons, as a consequence  of the equivalence
theorem, behave like scalar particles in the fundamental representation.
The structure of the Sudakov logarithms in these cases is significantly different
and we consider them separately.

\subsection{Transverse polarization}
\label{sec21} The transverse gauge bosons are the true vector
particles and the Born amplitude in this case is given by the
$t$-channel and $u$-channel fermion exchange diagrams, Fig.~\ref{fig1}$a$. It is
convenient to introduce the functions ${\cal Z}_\psi$ and ${\cal
Z}_A$ which describe the asymptotic dependence on the large momentum
transfer $Q$ of the fermion scattering amplitude in an external
singlet vector field and of the vector boson in an external scalar
singlet field, {\it i.e.} of the respective form factors in the  Euclidean region.
In  leading order in $M^2/Q^2$ these
functions are known to satisfy the following linear evolution equation  \cite{Mue,Col,Sen1}
\begin{eqnarray}
{\partial\over\partial\ln{Q^2}}{\cal Z}_i&=&
\left[\int_{M^2}^{Q^2}{\dd x\over x}\gm_i(\al(x))+\zeta_i(\al(Q^2))
+\xi_i(\al(M^2)) \right] {\cal Z}_i \,,
\label{evoleqz}
\end{eqnarray}
with the solution
\begin{eqnarray}
{\cal Z}_i&=&\exp \left\{\int_{M^2}^{Q^2}{\dd x\over x}
\left[\int_{M^2}^{x}{\dd x'\over x'}\gm_i(\al(x'))+\zeta_i(\al(x))
+\xi_i(\al(M^2))\right]\right\}
\,,
\label{evolsolz}
\end{eqnarray}
which satisfies the initial condition ${\cal Z}_i\big|_{Q^2=M^2}=1$.  Here the
perturbative functions $\gm_i(\al)$ {\it etc.} are given by the series
in the coupling constant $\alpha(\mu^2)$, {\it e.g.}
$\gm_i(\al)=\sum_{n=1}^\infty (\al/4\pi)^n\gm_i^{(n)} $. For the
amplitude of transverse boson production ${\cal A}_T$ let us
introduce the reduced amplitude $\tilde{\cal A}_T$ so that
\begin{equation} {\cal
  A}_T={\cal Z}_\psi{\cal Z}_A\tilde{\cal A}_T\,.
\label{ampred}
\end{equation}
Due to the factorization property of the Sudakov logarithms associated
with the {\it collinear} divergences of the massless theory
\cite{FreTay} the reduced amplitude satisfies the simple renormalization
group like equation \cite{Sen2,Ste,BotSte}
\begin{equation}
{\partial \over \partial \ln{Q^2}}\tilde {\cal A}_T=
\bfm{\chi}_T(\al(Q^2))\tilde {\cal A}_T \,,
\label{evoleqa}
\end{equation}
where $Q^2=-s$ and  $\bfm{\chi}_T$ is the soft anomalous dimension matrix
acting in the space of the isospin amplitudes.  The solution of the
above equation is given by the path-ordered exponent
\begin{equation}
\tilde {\cal A}_T=
{\rm P}\!\exp{\left[\int_{M^2}^{Q^2}
{\dd x\over x}\bfm{\chi}_T(\al(x))\right]}{{\cal A}_0}_T(\al(M^2))\,,
\label{evolsola}
\end{equation}
where ${{\cal A}_T}_0$ determines the initial conditions for the
evolution equation at $Q=M$.  By calculating the functions entering the
evolution equations order by order in $\al$ one gets the logarithmic
approximations for the amplitude.  For example, the LL approximation
includes all the terms of the form $\al^n{\cal L}^{2n}$ and is
determined by the one-loop coefficients $\gm_i^{(1)}$. The NLL
approximation includes all the terms of the form $\al^n{\cal L}^{2n-m}$
with $m=0,~1$ and requires the one-loop coefficients $\zeta_i^{(1)}$,
$\xi_i^{(1)}$, and $\bfm{\chi}^{(1)}$ as well as the one-loop running of
$\al$ in $\gm_i(\al)$, and so on.  To get the NNLL terms $\al^n{\cal
  L}^{2n-2}$ one needs in addition the two-loop coefficient
$\gm_i^{(2)}$,  the two-loop running of $\al$ in $\gm_i(\al)$ and
the one-loop contribution to ${{\cal A}_T}_0$.

The anomalous dimensions $\gm(\alpha)$, $\zeta(\alpha)$ and
$\bfm{\chi}(\al)$ are mass-independent and can be associated with
the infrared divergences of the massless (unbroken) theory.  From
the QCD result (see {\it e.g.} \cite{Cat} and references therein)
adopted for the specific case of $SU(2)$ gauge group, $n_f$ chiral
quarks, and one scalar in the fundamental representation we get
\begin{equation}
\gm_\psi^{(1)}=-{3/2}\,,\qquad \gm_\psi^{(2)}=-{65\over
3}+{\pi^2}+{5\over 6}n_f\,,\qquad \qquad\zeta_\psi^{(1)}={9\over
4}\,,\qquad \zeta_A^{(1)}=0\,, \label{1lgam}
\end{equation}
and $\gm_A^{(n)}=8\gm_\psi^{(n)}/3$. The matrix
$\bfm{\chi}_T^{(1)}$ can be extracted from the results of Refs.~
\cite{BFD,GloTej}. In the isospin basis
$(\bfm{\sigma}^a\bfm{\sigma}^b,~\bfm{\sigma}^b\bfm{\sigma}^a,
~\delta^{ab}\cdot\bfm{1})$ it takes the form
\begin{equation}
\bfm{\chi}^{(1)}_T=\pmatrix{ -2 (\ln(x_-)+i\pi) & 0 & \ln({x_+ \over
x_-})) \cr
          0 & -2 (\ln(x_+)+i\pi)  & \ln({x_- \over x_+}) \cr
    (\ln(x_+)+i\pi) & (\ln(x_-)+i\pi) & 0}\,,
\label{1lchit}
\end{equation}
where $x_\pm=(1\pm\cos\theta)/2$ and $\theta$ is the production
angle. Note that in this basis the Born amplitude up to a common factor
is given by the vector $(1/\xm,1/\xp,0)$.  At the same time the functions $\xi_i(\al)$
and ${\cal A}_T(\al)$ do depend on the infrared structure of the
model and require the calculation in the spontaneously broken phase.
For example $\xi_\psi^{(1)}=0$ \cite{KMPS} and from the result of
Ref.~\cite{Bee} we obtain $\xi_A^{(1)}=0$. To emulate the $e^+e^-\to
W^+_TW^-_T$ process one has to project the amplitude on the relevant
initial and final isospin states, which is straightforward.  The
Born cross section in the high energy limit reads
\begin{equation}
 {{\rm d} \sigma^B_T \over
  {\rm d} \Omega}={\alpha^2(M^2)\over 4s}{\xp(\xp^2+\xm^2)\over \xm}\,,
\label{bsigt}
\end{equation}
and  is peaked in the forward direction.
We define the perturbative expansion for the differential cross section
in the $\overline{\rm MS}$ renormalized coupling
constant $ \alpha\equiv\alpha(M^2)$  as follows
\begin{equation}
{{\rm d} \sigma \over {\rm d} \Omega}=
\left[1+\left( {\alpha \over 4\pi} \right)\dl^{(1)}
+\left( {\alpha\over 4\pi} \right)^2
\dl^{(2)}+\ldots\right]
{{\rm d} \sigma^B \over {\rm d} \Omega} \,.
\label{sersig}
\end{equation}
Expanding the Sudakov
exponents to NNLL order for the one- and two-loop corrections we get
\begin{eqnarray}
\dl_T^{(1)}&=& \left(\gamma^{(1)}_\psi+\gamma^{(1)}_A\right){\cal
L}^2(s)+2\left[\zeta^{(1)}_\psi+\zeta^{(1)}_A+\xi^{(1)}_\psi+\xi^{(1)}_A
+t_{11}^{(1)}+t_{31}^{(1)}+ {x_-\over x_+}
\left(t_{12}^{(1)}+t_{32}^{(1)}\right)\right]
\nn\\
&& \times{\cal L}(s) + \dl_{0T}^{(1)} \label{1lsigta}
\\
\dl_T^{(2)}&=& {1\over
2}\left(\gamma^{(1)}_\psi+\gamma^{(1)}_A\right)^2{\cal
L}^4(s)+2\Bigg[\zeta^{(1)}_\psi+\zeta^{(1)}_A+\xi^{(1)}_\psi+\xi^{(1)}_A
+t_{11}^{(1)}+t_{31}^{(1)}+ {x_-\over x_+}
\left(t_{12}^{(1)}+t_{32}^{(1)}\right)
\nn\\
&& -{1\over
6}\beta_0\Bigg]\left(\gamma^{(1)}_\psi+\gamma^{(1)}_A\right){\cal
L}^3(s) + \Bigg[\gamma^{(2)}_\psi+\gamma^{(2)}_A
+2\left(\zeta^{(1)}_\psi+\zeta^{(1)}_A+\xi^{(1)}_\psi+\xi^{(1)}_A
\right)\Bigg(\zeta^{(1)}_\psi+\zeta^{(1)}_A
\nn\\
&&\left.+\xi^{(1)}_\psi+\xi^{(1)}_A+2\left(t_{11}^{(1)}+t_{31}^{(1)}+ {x_-\over x_+}
\left(t_{12}^{(1)}+t_{32}^{(1)}\right)\right)\right)
-\beta_0\left(t_{11}^{(1)}+t_{31}^{(1)}+ {x_-\over x_+}
\left(t_{12}^{(1)}+t_{32}^{(1)}\right)\right.
\nn\\
&& \zeta^{(1)}_\psi+\zeta^{(1)}_A\Bigg)+
{t_{11}^{(1)}}^2+t_{12}^{(1)}t_{21}^{(1)}+t_{11}^{(1)}t_{31}^{(1)}+
t_{13}^{(1)}t_{31}^{(1)}+t_{21}^{(1)}t_{32}^{(1)}+t_{31}^{(1)}t_{33}^{(1)}
+ {x_-\over
x_+}\nn\\
&&
\times\left(t_{12}^{(1)}\left(t_{11}^{(1)}+t_{22}^{(1)}+t_{31}^{(1)}\right)
+t_{32}^{(1)}\left(t_{13}^{(1)}+t_{22}^{(1)}+t_{33}^{(1)}\right)\right)
+\left(t_{11}^{(1)}+t_{31}^{(1)}+ {x_-\over x_+}
\left(t_{12}^{(1)}+t_{32}^{(1)}\right)\right)^2\nn\\
&&
+ \dl_{0T}^{(1)}\left(\gamma^{(1)}_\psi+\gamma^{(1)}_A\right)\Bigg]
{\cal L}^2(s) \label{2lsigta}
\end{eqnarray}
where $\beta_0=43/6-n_f/3$ is the one-loop beta-function,
$t_{ij}^{(1)}\equiv {\rm Re}[\bfm{\chi}^{(1)}_T]_{ij}$, ${\cal
L}(s)\equiv\ln(s/M^2)$, and $\dl_{0T}^{(1)}$ is the nonlogarithmic
part of the one-loop corrections which can be extracted from the
result of Ref.~\cite{Bee}.  For the Higgs boson mass $M_H=M$ it
takes a simple form
\begin{eqnarray}
\dl_{0T}^{(1)}&=&\left({5+3\xm \over 2(\xm^2+\xp^2)}
-{5 \over \xp}\right)\ln^2(\xm)
+{3\xm \over \xm^2+\xp^2}\ln(\xp^2)+{4 \over \xp}\ln(\xm)\ln(\xp)
\nn\\
&&+{9-19\xm \over 2(\xm^2+\xp^2)}\ln(\xm) -{5\xp \over
2(\xm^2+\xp^2)} -{7\pi^2 \over 18}-{13\pi \over 3 \sqrt{3}}+{455
\over 36}\,-{10 \over 9}n_f. \label{del0t}
\end{eqnarray}
Substituting the values of the coefficients for $n_f=12$ we obtain
\begin{eqnarray}
\dl_T^{(1)}&=&-{11 \over 2} {\cal L}^2(s)
+\left[\left(-8+{4 \xm \over \xp}\right)\ln(\xm)+4 \ln(\xp)
+{9\over 2}\right]{\cal L}(s)
\nn\\
&&+\left({5+3\xm \over 2(\xm^2+\xp^2)}
-{5 \over \xp}\right)\ln^2(\xm)
+{3\xm \over \xm^2+\xp^2}\ln(\xp^2)+{4 \over \xp}\ln(\xm)\ln(\xp)
\nn\\
&&+{9-19\xm \over 2(\xm^2+\xp^2)}\ln(\xm)
-{5\xp \over 2(\xm^2+\xp^2)}
-{7\pi^2 \over 18}-{13\pi \over 3 \sqrt{3}}-{25 \over 36}\,,
\label{1lsigt}\\
\dl_T^{(2)}&=&{121 \over 8} {\cal L}^4(s)
+\left[\left(44-{22\xm \over \xp}\right)\ln(\xm)-22\ln(\xp)
-{341 \over 18}\right]{\cal L}^3(s)
\nn\\
&&+\left[\left(32+{4\xm^2 \over \xp^2}
-{55+33\xm \over 4 (\xm^2+\xp^2)}
+{55-40\xm \over 2\xp}\right)\ln^2(\xm)
+\left(8-{33\xm \over 2(\xm^2+\xp^2)}\right)\ln^2(\xp)
\right.
\nn\\
&&\left.-\left(28
+{22 -4\xm\over \xp}\right)\ln(\xm)\ln(\xp)
+\left(-{70 \over 3}+{35\xm \over 3\xp}
-{99-209\xm \over 4(\xm^2+\xp^2)}\right)\ln(\xm)
\right.
\nn\\
&&\left.+{35 \over 3}\ln(\xp)
+{55\xp \over 4(\xm^2+\xp^2)}+{209\pi^2 \over 36}
+{143\pi \over 6\sqrt{3}}-{863 \over 24}\right]{\cal L}^2(s)\,.
\label{2lsigt}
\end{eqnarray}
Note that in contrast to the four-fermion processes, the cross
section of the gauge boson production depends on the Higgs boson
mass already in  the NNLL approximation.

\subsection{Longitudinal polarization}
\label{sec22} The equivalence theorem relates the amplitude of the
longitudinal gauge boson production $e^+e^-\to W^+_LW^-_L$ to the
production of  the Goldstone bosons $e^+e^-\to\phi^+\phi^-$. The
Born amplitude is now  given by the $s$-channel annihilation
diagram, Fig.~\ref{fig1}$b$. The analysis of the high energy
asymptotic for the longitudinal gauge boson production goes along
the line described in the previous section and is very similar to
the one for fermion pair production \cite{KMPS}. Instead of ${\cal
Z}_A$ one should use the function ${\cal Z}_\phi$ which correspond
to the scalar particle scattering in an external singlet vector
field. The necessary parameters of the evolution equation read
$\gm_\phi^{(n)}=\gm_\psi^{(n)}$, $\zeta_\phi^{(1)}=3$, and from the
result of Ref.~\cite{Bee} we get $\xi_\phi^{(1)}=0$. The structure
of the reduced amplitude is also different.  In the isospin basis
$(\bfm{\sigma}^a\otimes\bfm{\sigma}^a,\bfm{1}\otimes\bfm{1})$ the
one-loop matrix of the soft anomalous dimensions takes the form
\begin{equation}
\bfm{\chi}^{(1)}_L= \pmatrix{
-4\left(\ln\left({x_+}\right)+i\pi\right)+
2\ln\left({x_+\over x_-}\right) \,\,&
4\ln\left({x_+\over x_-}\right) \cr
{3\over 4}\ln\left({x_+\over x_-}\right) \label{china} &
0}\,.
\label{1lchil}
\end{equation}
Note that in this basis the Born amplitude is given by the vector $(1,0)$.
The corresponding Born cross section  reads
\begin{equation}
 {{\rm d} \sigma^B_L \over
  {\rm d} \Omega}={\alpha^2(s)\over 4s}{\xp\xm\over 4}
\,.
\label{bsigl}
\end{equation}
It has a maximum at $\theta=90^o$.  We proceed as in the case of
transverse polarization and obtain the one- and two-loop NNLL
corrections to the differential cross section
\begin{eqnarray}
\dl_L^{(1)}&=& \left(\gamma^{(1)}_\psi+\gamma^{(1)}_\phi\right){\cal
L}^2(s)+2\left[\zeta^{(1)}_\psi+\zeta^{(1)}_\phi+\xi^{(1)}_\psi+\xi^{(1)}_\phi
+l_{11}^{(1)}+4l_{21}^{(1)}\right]{\cal L}(s) + \dl_{0L}^{(1)} \label{1lsigla}
\\
\dl_L^{(2)}&=& {1\over
2}\left(\gamma^{(1)}_\psi+\gamma^{(1)}_\phi\right)^2{\cal
L}^4(s)+2\left[\zeta^{(1)}_\psi+\zeta^{(1)}_\phi+\xi^{(1)}_\psi+\xi^{(1)}_\phi
+l_{11}^{(1)}+4l_{21}^{(1)}-{1\over
6}\beta_0\right]\left(\gamma^{(1)}_\psi+\gamma^{(1)}_\phi\right)
\nn\\
&& \times{\cal
L}^3(s) + \Bigg[\gamma^{(2)}_\psi+\gamma^{(2)}_\phi
+2\left(\zeta^{(1)}_\psi+\zeta^{(1)}_\phi+\xi^{(1)}_\psi+\xi^{(1)}_\phi
\right)\left(\zeta^{(1)}_\psi+\zeta^{(1)}_\phi+\xi^{(1)}_\psi+\xi^{(1)}_\phi
\right.\nn\\
&&
\left.+2l_{11}^{(1)}+8l_{21}^{(1)}\right)-\beta_0\left(l_{11}^{(1)}+4l_{21}^{(1)}+\zeta^{(1)}_\psi+\zeta^{(1)}_\phi\right)
+{l_{11}^{(1)}}^2+l_{21}^{(1)}\left(4l_{11}^{(1)}+l_{12}^{(1)}+4l_{22}^{(1)}\right)
\nn\\
&&
+\left(l_{11}^{(1)}+4l_{21}^{(1)}\right)^2+ \dl_{0L}^{(1)}\left(\gamma^{(1)}_\psi+\gamma^{(1)}_\phi\right)\Bigg]
{\cal L}^2(s) \nn\\
&&\label{2lsigla}
\end{eqnarray}
where  $l_{ij}^{(1)}\equiv {\rm Re}[\bfm{\chi}^{(1)}_L]_{ij}$. From
the result of Ref.~\cite{Bee}  for $M_H=M$ we obtain\footnote{The
equivalence theorem holds up to the field renormalization factor
(see {\it e.g.} Ref.~\cite{Bag}) which affects the initial
conditions for the evolution equation. Thus one has to use the
explicit result for the longitudinal W-boson production rather than
the equivalence theorem to get the momentum and angular independent
term in $\dl_{0L}^{(1)}$.}
\begin{eqnarray}
\dl_{0L}^{(1)}&=&-{5 \over 2\xp}\ln^2(\xm) +{1 \over 2\xm}\ln^2(\xp)
-{7\pi^2 \over 3}+{32\pi \over 3 \sqrt{3}}-{25 \over 36}-{10\over
9}n_f\,. \label{del0l}
\end{eqnarray}
For $n_f=12$ this gives
\begin{eqnarray}
\dl_L^{(1)}&=&-3{\cal L}^2(s)
+\left[-10\ln(\xm)+2\ln(\xp)+{21\over 2}\right]{\cal L}^2(s)
\nn\\
&&-{5 \over 2\xp}\ln^2(\xm)
+{1 \over 2\xm}\ln^2(\xp)
-{7\pi^2 \over 3}+{32\pi \over 3 \sqrt{3}}-{505 \over 36}\,,
\label{1lsigl} \\
\dl_L^{(2)}&=&{9 \over 2} {\cal L}^2(s)
+\left[30\ln(\xm)-6\ln(\xp)-{85\over 3}\right]{\cal L}^3(s)
\nn\\
&&+\left[\left(38+{15 \over 2\xp}\right)\ln^2(\xm)
+\left(2-{3 \over 2\xm}\right)\ln^2(\xp)
-8\ln(\xm)\ln(\xp)
\right.
\nn\\
&&\left.-{535 \over 6}\ln(\xm)+{107 \over 6}\ln(\xp)
+9\pi^2 -{32\pi \over \sqrt{3}}+{229 \over 4}\right]{\cal L}^2(s)\,.
\label{2lsigl}
\end{eqnarray}

\section{$W$-pair production in $e^+e^-$ annihilation}
\label{sec3}
In the standard electroweak model the perturbative expansion involves
the $SU_L(2)$ coupling constant $\al_{ew}$ and the $U(1)$ hypercharge
coupling constant $\al_Y$.  We eliminate the latter by means of the
relation $\al_Y=\tan^2\theta_W\,\al_{ew}$, where $\theta_W$ is the weak mixing
angle, and consider the one-parameter series for the cross section in
$\al_{ew}$ of the form of Eq.~(\ref{sersig}).  In the high energy limit the
transverse gauge bosons are produced only in annihilation of the
left-handed electron-positron pair.  The corresponding Born cross
section is given by Eq.~(\ref{bsigt}) with $\al$ replaced by
$\al_{ew}$.  The longitudinal gauge bosons
can be produced in the annihilation of the electron-positron pair of
both chiralities.  In the case of the left-handed initial state fermions
the Born cross section gets the contribution from the $SU_L(2)$ and the
hypercharge virtual gauge bosons and reads
\begin{equation}
 {{\rm d} \sigma^B_{-L} \over
  {\rm d} \Omega}={1\over \cos^4\theta_W}{\alpha_{ew}^2(s)\over 4s}{\xp\xm\over 4}
\,.
\label{bsiglm}
\end{equation}
For the right-handed initial state fermions the Born cross section
is saturated by the hypercharge gauge boson
\begin{equation}
{{\rm d} \sigma^B_{+L}\over {\rm d} \Omega}=4\sin^4\theta_W\,{{\rm d}
\sigma^B_{-L}\over{\rm d} \Omega}\,.
\label{bsiglp}
\end{equation}
The analysis of the radiative corrections  in the Standard Model
is complicated by the presence of the mass gap and mixing in the gauge sector
and by the large top quark Yukawa coupling.  In next section we use the
method of Ref.~\cite{Fad,JKPS} to separate the electroweak and QED
Sudakov logarithms. In Sect.~\ref{sec32} we extend the evolution
equation approach to the Yukawa enhanced contributions.  In
Sect.~\ref{sec33} we present the final numerical results for the
two-loop NNLL corrections to the differential cross sections.

\subsection{Separating QED Sudakov logarithms}
\label{sec31} The electroweak Standard Model with the spontaneously
broken $SU_L(2)\times U(1)$ gauge group involves both the massive
$W$ and $Z$ bosons and the massless photon. The corrections to the
fully exclusive cross sections due to the virtual photon exchange
are infrared divergent and should be combined with soft real
photon emission to obtain infrared finite physical observables. We
regularize the infrared divergences by giving the photon a small
mass $\lm$. Thus besides the electroweak Sudakov logarithms discussed
above the radiative corrections contain the QED Sudakov logarithms
of the form $\ln(Q^2/\lm^2)$. To disentangle the electroweak and QED
logarithms we use the approach of Ref.~\cite{Fad,KMPS,JKPS}. While
the dependence of the amplitudes on the large momentum transfer is
governed by the {\it hard} evolution equations ({\it  c.f.}
Eqs.~(\ref{evoleqz},~\ref{evoleqa})), their dependence on the photon
mass is governed by the {\it infrared} evolution equations
\cite{Fad}.  In the limit $\lm^2\ll M_W^2,~m_t^2\ll Q^2$ the infrared
evolution equations in the full theory are the same as in QED and the
solution to the NNLL accuracy in the massless fermion approximation
$m_f=0~(f\ne t)$ is given by the factor
\begin{eqnarray}
{\cal U}&=&U_0(\al_e)\exp{\left\{ -{\alpha_e(\lm^2)\over
4\pi}\left[\left(2 -\left( {290\over 27}+{40\over
9}\ln\left({x_+\over x_-}\right)\right){\alpha_e\over \pi}
\right)\ln^2\left({Q^2\over \lm^2}\right)\right.\right.}
\nn\\
& &-\left(3+ 4\ln\left({x_+\over x_-}\right)\right)
\ln\left({Q^2\over\lm^2}\right) +{40\over 27}{\alpha_e\over \pi}
\ln^3\left({Q^2\over\lm^2}\right)-
\left(\ln\left({M_W^2\over\lm^2}\right)-1\right)^2\Bigg]
\nn\\
&&+{\cal O}(\al_e^3)\Bigg\},
\label{QED}
\end{eqnarray}
where $\al_e$ is the $\overline{\rm MS}$ QED coupling constant. The NNLL approximation
for ${\cal U}$ can be obtained from the result for the
fermion-antifermion production  \cite{KMPS} by proper modification
of the QED anomalous dimensions. It is convenient to normalize
the QED factor so that ${\cal U}(\al_e)\big|_{s=\lm^2=M_W^2}=1$.
In order to cancel  the singular
dependence on the photon mass, the QED Sudakov
exponent~(\ref{QED}) should be combined with the the real photon
emission, which is also of pure QED nature if the  energy of emitted
photons  is much smaller than $M_{W}$.

Two sets of equations completely determine the dependence of the
amplitudes on two dimensionless variables $Q/M_W$ and $Q/\lm$ up to
the initial conditions which are fixed through the matching
procedure. To get the purely weak logarithms one subtracts the QED
exponent~(\ref{QED}) from the
exponent given by the solution of the hard evolution equation. This
can naturally be formulated in terms of the functions parameterizing
the solution. The functions  $\gm$, $\zeta$, and
$\bfm{\chi}$ are  mass-independent. Therefore the anomalous dimensions
parameterizing the purely  weak logarithms can be obtained by
subtracting the QED contribution  from the result of the unbroken symmetry
phase calculation to all orders in the coupling constants. In the order of
interest they can be found in or easily derived from the result of
Ref.~\cite{KMPS}. For example, we have
\begin{eqnarray}
\gm_A^{(2)}&=&\gm_A^{(2)}|_{SU(2)}-{800\over 27}\sin^2\theta_W\,,
\\
\gm_\phi^{(2)}&=&\gm_\phi^{(2)}|_{SU(2)}+{52\over
9} \tan^2\theta_W-{800\over
27}\sin^2\theta_W\,,\label{g2sub}
\end{eqnarray}
and so on. Here the  $SU(2)$ contributions are given by the results
of Sect.~\ref{sec2} with $n_f=12$.
The only new ingredient in comparison with the light fermion pair production
\cite{KMPS} is the effect of the  large Yukawa coupling of the
third generation quarks on the  longitudinal gauge boson production
which is considered in the next section.

On the other hand the  functions
$\xi$ and ${\cal A}_0$ are infrared sensitive and require the use of
the  true mass eigenstates of the Standard Model in the perturbative
calculation. In the NNLL approximation one needs the one-loop
contribution to these quantities which can be extracted from the
result of Ref.~\cite{Bee}. For example, for the left-handed initial state fermions we find
\begin{eqnarray}
\xi_\psi^{(1)}+\xi_A^{(1)}&=&{1-4\cos^2\theta_W+8\cos^4\theta_W\over
2\cos^2\theta_W}\ln\left({M_Z^2\over M_W^2}\right)\,,
\nn\\
\xi_\psi^{(1)}+\xi_\phi^{(1)}&=&{(1-2\cos^2\theta_W)^2\over
\cos^2\theta_W}\ln\left({M_Z^2\over M_W^2}\right)\,. \label{xi1ew}
\end{eqnarray}
At the same time the ${\cal A}_0^{(1)}$ term results in the one-loop
nonlogarithmic contribution to the cross section. The corresponding
expression directly follows from Ref.~\cite{Bee} and is rather
lengthy so we do not give it explicitly. Note that in
Ref.~\cite{Bee} the result is presented in the on-shell
renormalization scheme and in the limit $\lm\ll m_e$. We convert
it to $\overline{\rm MS}$ scheme and to the massless
electron approximation using the formulae of Refs.~\cite{FKS,Pen}.

\subsection{Top quark Yukawa coupling effects}
\label{sec32} The large Yukawa coupling of the third generation
quarks to the scalar (Higgs and Goldstone) bosons results in
specific logarithmic corrections proportional to  $m_t^2/M_W^2$. The
high energy evolution of the form factors in  a theory with Yukawa
interaction is completely analogous to the one of $\phi^3$ scalar
theory in six dimensions, see the second paper of Ref.~\cite{Col}.
The structure of factorization and  evolution equations is much
simpler than in a gauge theory because Yukawa interaction itself
does not contribute to the anomalous dimension $\gamma_i(\al)$ and
results only in single logarithmic corrections completely determined
by the ultraviolet field renormalization of the external on-shell
particles. These corrections can be taken into account through the
modification of the evolution equations for the corresponding $\cal
Z$-functions. The analysis is straightforward but complicated
because the Yukawa interaction mixes  evolution of the quark and
scalar boson form factors and in general does not commute with the
$SU(2)$ and hypercharge couplings. However, due to the factorization
of the double Sudakov logarithms, the Yukawa enhanced contribution
to NLL approximation is given simply by the product of the one-loop
Yukawa corrections and the double logarithmic exponent as observed
in Ref.~\cite{Mel2}. The structure of the NNLL contribution is much
more complicated and we restrict the analysis to a simplified model
with $\sin\theta_W=0$, {\it i.e.} with no hypercharge interaction.
Let us introduce the following five-component vector in the space of
$\cal Z$-functions $\bfm{\cal Z}=({\cal Z}_\phi,~{\cal
Z}_\chi,~{\cal Z}_{b-},~{\cal Z}_{t-},~{\cal Z}_{t+})$, where the
subscript $+$ $(-)$ stand for the right (left) quark  fields and
${\cal Z}_\chi$ corresponds to the transition of the Higgs boson
into the neutral Goldstone boson in the external singlet vector
field. The evolution equation for this vector takes the form
\begin{figure}[t]
\begin{center}
\hspace*{10mm}
\begin{picture}(100,60)(0,0)
\SetScale{0.8} \SetWidth{1} \ArrowLine(0,30)(50,60)
\ArrowLine(50,0)(0,30) \ArrowLine(50,60)(50,0)
\DashLine(50,60)(100,60){5} \DashLine(50,00)(100,00){5}
\CBoxc(0,30)(6,6){Black}{Black}
\end{picture}
\hspace*{10mm}
\begin{picture}(100,60)(0,0)
\SetScale{0.8} \SetWidth{1} \DashLine(0,30)(50,60){5}
\DashLine(50,0)(0,30){5} \ArrowLine(50,60)(50,0)
\ArrowLine(50,60)(100,60) \ArrowLine(50,00)(100,00)
\CBoxc(0,30)(6,6){Black}{Black}
\end{picture}
\hspace*{10mm}
\begin{picture}(100,60)(0,0)
\SetScale{0.8} \SetWidth{1} \ArrowLine(0,30)(50,60)
\ArrowLine(50,0)(0,30) \DashLine(50,60)(50,0){5}
\ArrowLine(50,60)(100,60) \ArrowLine(100,00)(50,00)
\CBoxc(0,30)(6,6){Black}{Black}
\end{picture}
\end{center}
\caption{\label{fig2} \small  The one-loop diagrams contributing to
the anomalous dimension matrix  $\bfm{\zeta}$. The arrow lines
correspond to the third generation quarks. The dashed lines
correspond to the Higgs, neutral or charged Goldstone bosons. The
black square represent an external singlet vector field}
\end{figure}
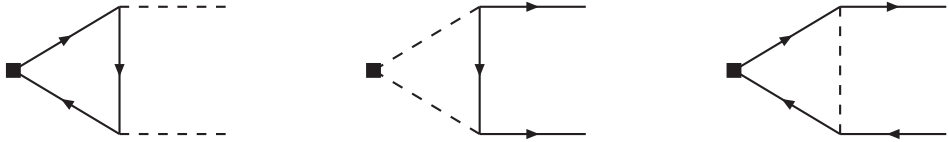
\begin{equation}
{\partial\over\partial\ln{Q^2}}\bfm{\cal Z}=
\Bigg[\int_{M_W^2}^{Q^2}{\dd x\over
x}\bfm{\gm}(\al_{ew}(x))+\bfm{\zeta}(\al_{ew}(Q^2),\alpha_{Y\!uk}(Q^2))
+\bfm{\xi}(\al_{ew}(M_W^2)) \Bigg] \bfm{\cal Z} \,,
\label{evoleqzvec}
\end{equation}
with the solution
\begin{equation}
\bfm{\cal Z}={\rm P}\!\exp \left\{\int_{M_W^2}^{Q^2}{\dd x\over x}
\left[\int_{M_W^2}^{x}{\dd x'\over
x'}\bfm{\gm}(\al_{ew}(x'))+\bfm{\zeta}(\al_{ew}(x),\alpha_{Y\!uk}(x))
+\bfm{\xi}(\al_{ew}(M_W^2))\right]\right\}\bfm{\cal Z}_0 \,,
\label{evolsolzvec}
\end{equation}
where  $\bfm\gamma^{(1)}=(-3/2)\cdot\bfm{1}$ and $\bfm{\xi}=0$. The
anomalous dimension matrix $\bfm{\zeta}$ includes all the dependence
on the Yukawa coupling $\alpha_{Y\!uk}$. We eliminate the latter by
means of the relation $\alpha_{Y\!uk}={m_t^2\over 2M_W^2}\al_{ew}$,
and consider  the one-parameter series for the anomalous dimension
in $\al_{ew}$. The one-loop coefficient reads
\begin{equation}
\zeta^{(1)}={1\over 4}\pmatrix{ 12 & 0 & 0 & 0 & 0\cr
           0 & 12 & 0 & 0 &  0 \cr
           0& 0 & {9} & 0 & 0\cr
           0 & 0& 0 & {9} &  0 \cr
           0& 0 & 0&0 &{0} \cr
           }\,+{m_t^2\over 4M_W^2}\pmatrix{ 0 & 0 & {6} & 0 & -{6}\cr
           0 & 0 & 0 & {6} & -{6}\cr
           {1}& 0 & 0 & 0 & -{1}\cr
           0 & {1}& 0 & 0& -{1} \cr
           -{1} & -{1}  & -{1} &-{1}  &0 \cr
           }\,,
\label{1lzetayuk}
\end{equation}
where the first term representing the  pure $SU_L(2)$ contribution
follows from the result of Sect.~\ref{sec21} and the second term
represents the Yukawa contribution. It can be extracted from the
known one-loop result (see {\it e.g.} Ref.~\cite{Mel2,Bec}). The
relevant diagrams are given in Fig.~\ref{fig2}. Note that instead of
the ${\cal Z}$-functions associated with the form factors one can
directly consider the ultraviolet  field renormalization.  In this
case the non-diagonal form of the anomalous dimension matrix is due
to the mixing of the bilinear quark and scalar boson operators,
which is specific for Yukawa interaction and is absent in a gauge
theory.

 The first two diagrams  correspond to the mixing of the quark
and the scalar boson form factors. Moreover the Yukawa coupling
changes quark chirality and/or flavor and the last diagram in
Fig.~\ref{fig2} corresponds to the pure mixing of ${\cal Z}_{b-}$,
${\cal Z}_{t-}$ and ${\cal Z}_{t+}$ functions. As a consequence, all
the diagonal matrix elements in the second term of
Eq.~(\ref{1lzetayuk}) vanish.

The proper initial condition for the evolution equation which
corresponds to the Born amplitudes of the quark and scalar boson
production in $e^+e^-$ annihilation is given by the vector
$\bfm{\cal Z}_0=(1,-1,-1,1,0)$. In NNLL approximation one needs also
the one-loop running of the Yukawa coupling with the corresponding
beta-function $\beta_0^{Y\!uk}={9\over 4}-{3m_t^2\over 4M_W^2}$. By
expanding the solution for  the component ${\cal Z}_\phi$ we obtain
the two-loop corrections enhanced by the second or fourth power of
the top quark mass. Note that in the production amplitude one  has
to take into account also the interference between the one-loop
Yukawa contribution to ${\cal Z}_\phi$  and the one-loop logarithmic
term in the reduced amplitude and the electron ${\cal Z}_\psi$
function. The two-loop NNLL Yukawa enhanced contribution to the
cross section reads
\begin{equation}
\delta^{(2)}_{NNLL}\bigg|_{Y\!uk}=\left[{3\over 2}{m_t^4\over
M_W^4}+{m_t^2\over M_W^2}\left(30 \ln(\xm)-6\ln(\xp)-27\right)
\right] {\cal L}^2(s)\,. \label{2ldelyuk}
\end{equation}
This expression  approximates the full result up to the terms
suppressed by  $\sin^2\theta_W\sim 0.2$.

\begin{figure}
\begin{center}
\begin{tabular}{cc}
\hspace*{-25pt}\epsfig{figure=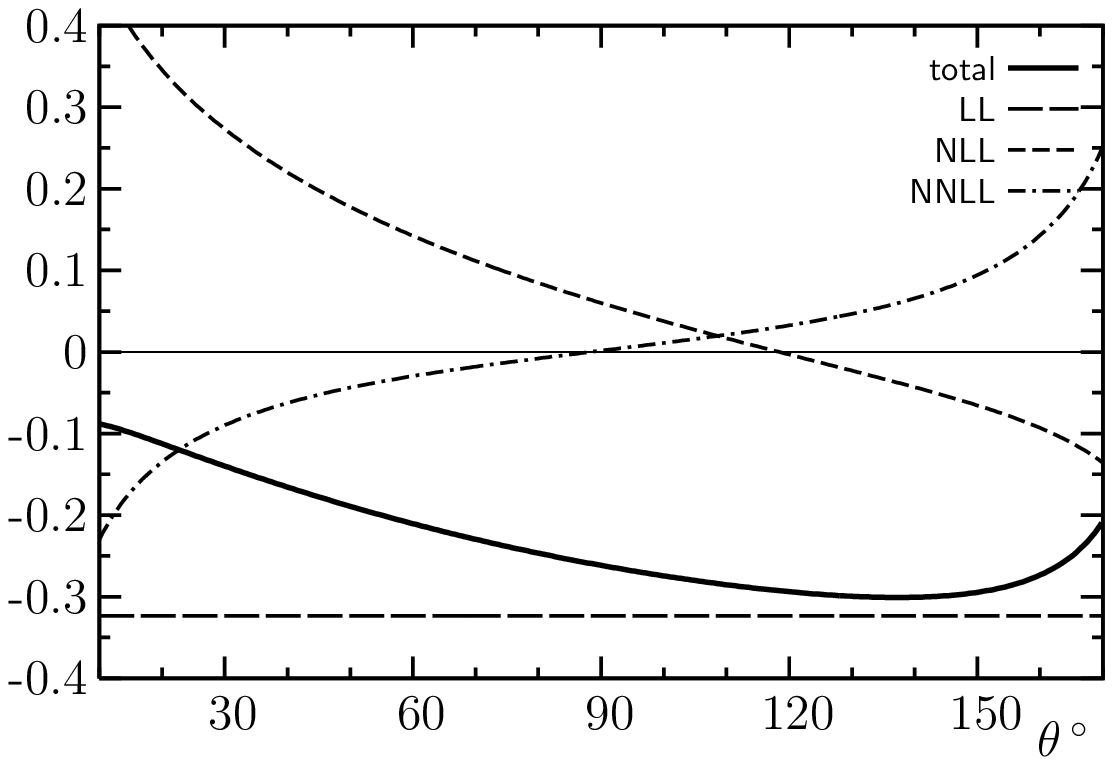,height=4.5cm}&
\hspace*{25pt}\epsfig{figure=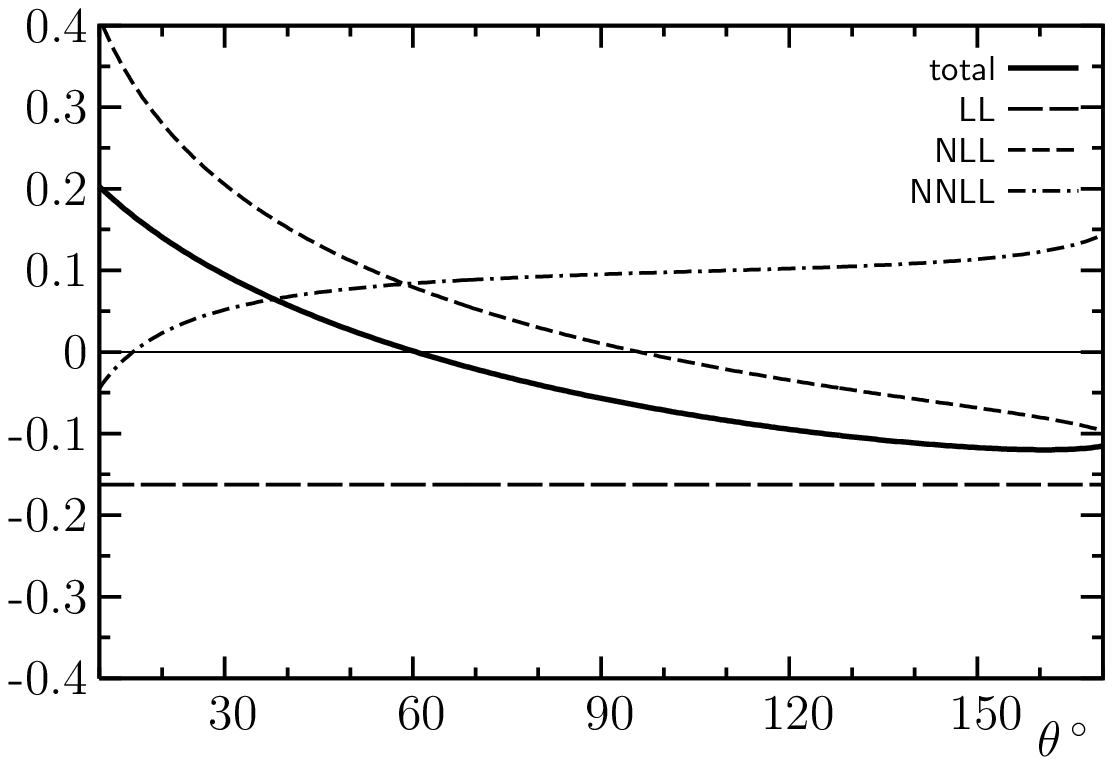,height=4.5cm}\\
\hspace*{-2pt}$(a)$ & \hspace*{47pt}$(b)$
\end{tabular}
\caption{\label{fig3} \small
  The one-loop logarithmic corrections to the differential cross section
  relative to the Born approximation at $\sqrt{s}=1$~TeV as functions of
  the production angle for $(a)$ transverse and $(b)$ longitudinal
  polarization of the gauge bosons.}
\end{center}
\end{figure}

\subsection{Numerical results}
\label{sec33} 
\begin{figure}
\begin{center}
\begin{tabular}{cc}
\hspace*{-25pt}\epsfig{figure=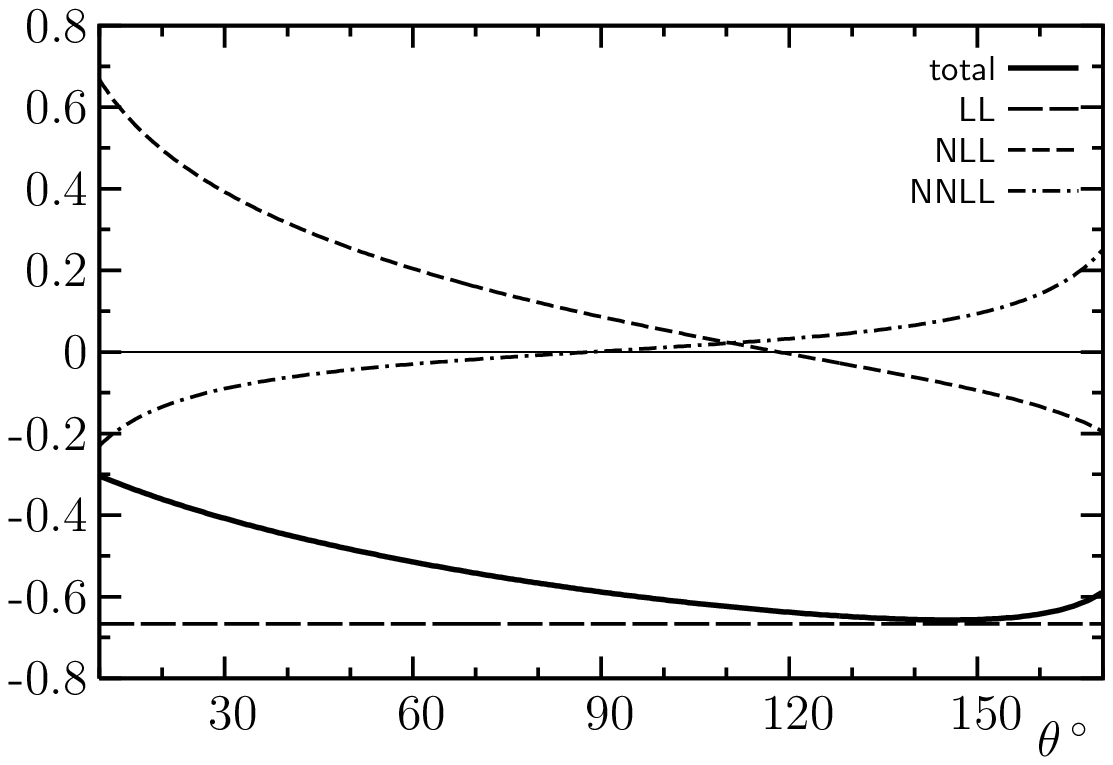,height=4.5cm}&
\hspace*{25pt}\epsfig{figure=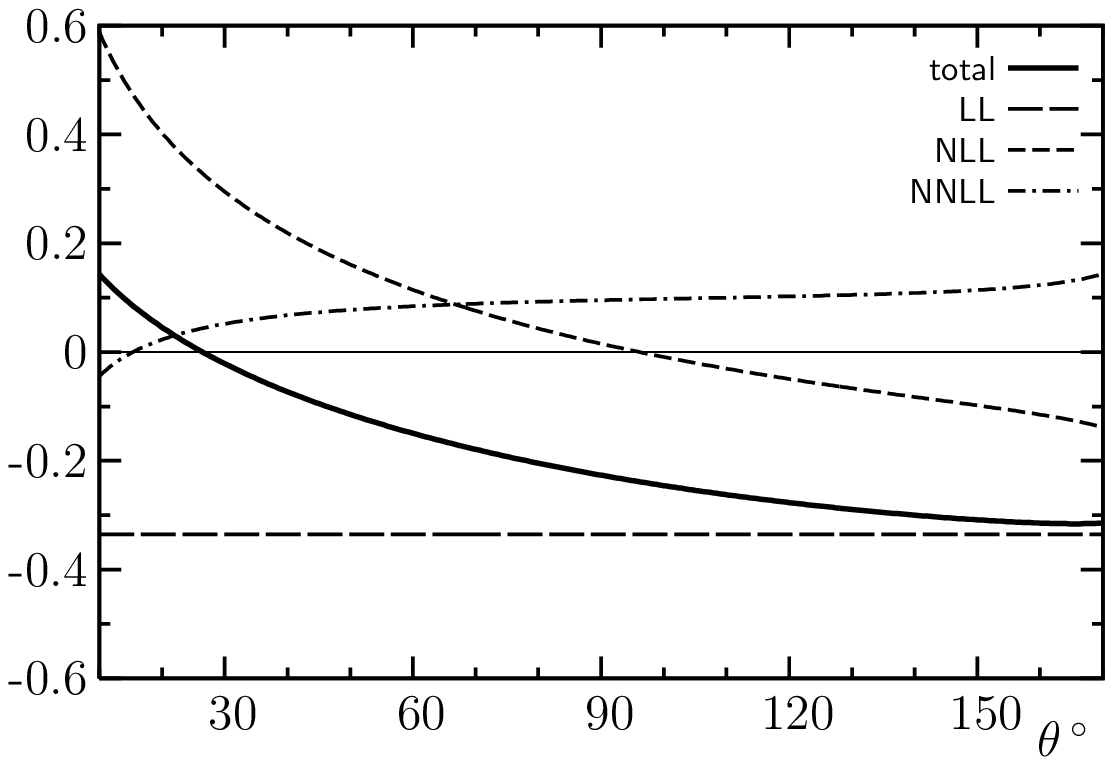,height=4.5cm}\\
\hspace*{-2pt}$(a)$ & \hspace*{47pt}$(b)$
\end{tabular}
\caption{\label{fig4} \small
  The same as Fig.~\ref{fig3} but for $\sqrt{s}=3$~TeV.}
\end{center}
\end{figure}

\begin{figure}
\begin{center}
\begin{tabular}{cc}
\hspace*{-25pt}\epsfig{figure=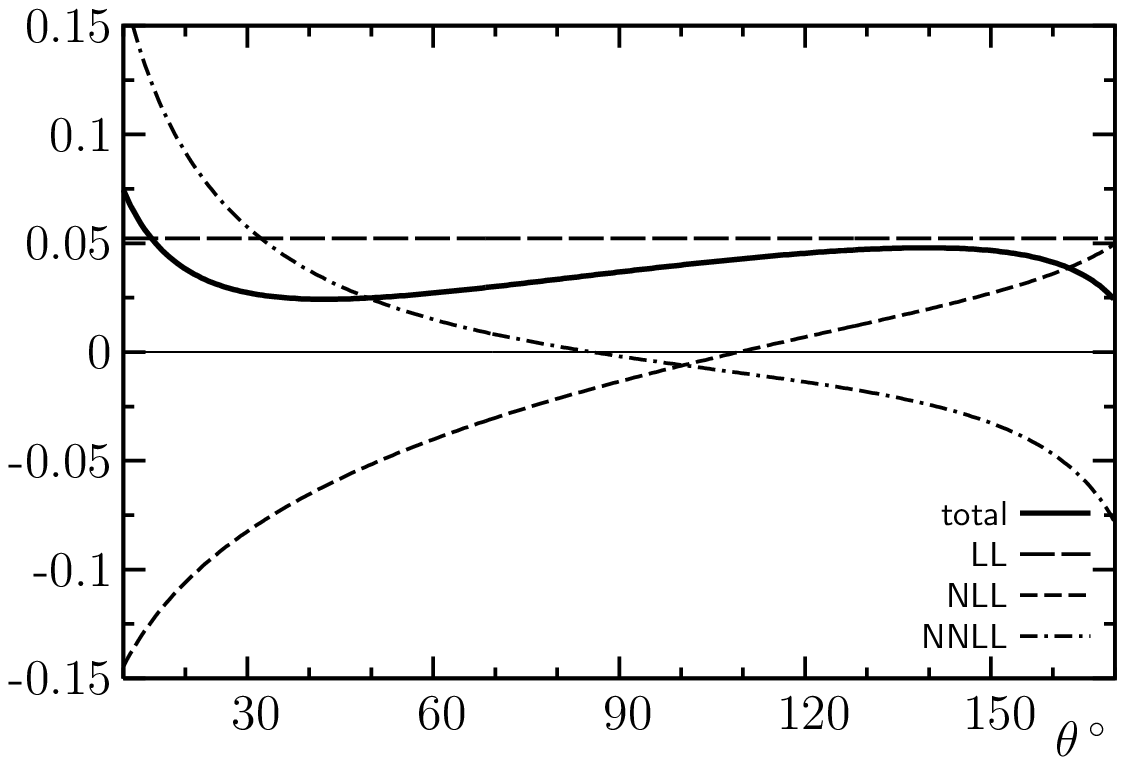,height=4.5cm}&
\hspace*{25pt}\epsfig{figure=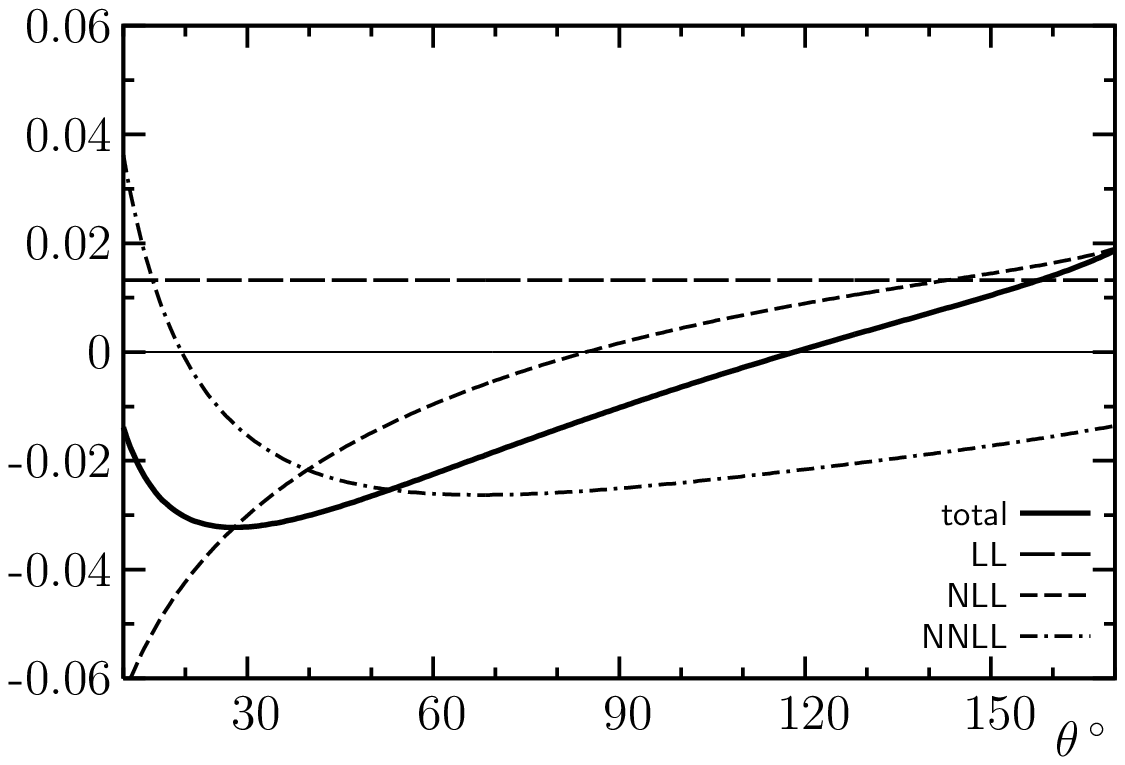,height=4.5cm}\\
\hspace*{-2pt}$(a)$ & \hspace*{47pt}$(b)$
\end{tabular}
\caption{\label{fig5} \small
  The two-loop logarithmic corrections to the differential cross section
  relative to the Born approximation at $\sqrt{s}=1$~TeV as functions of
  the production angle for $(a)$ transverse and $(b)$ longitudinal
  polarization of the gauge bosons.}
\end{center}
\end{figure}

\begin{figure}
\begin{center}
\begin{tabular}{cc}
\hspace*{-25pt}\epsfig{figure=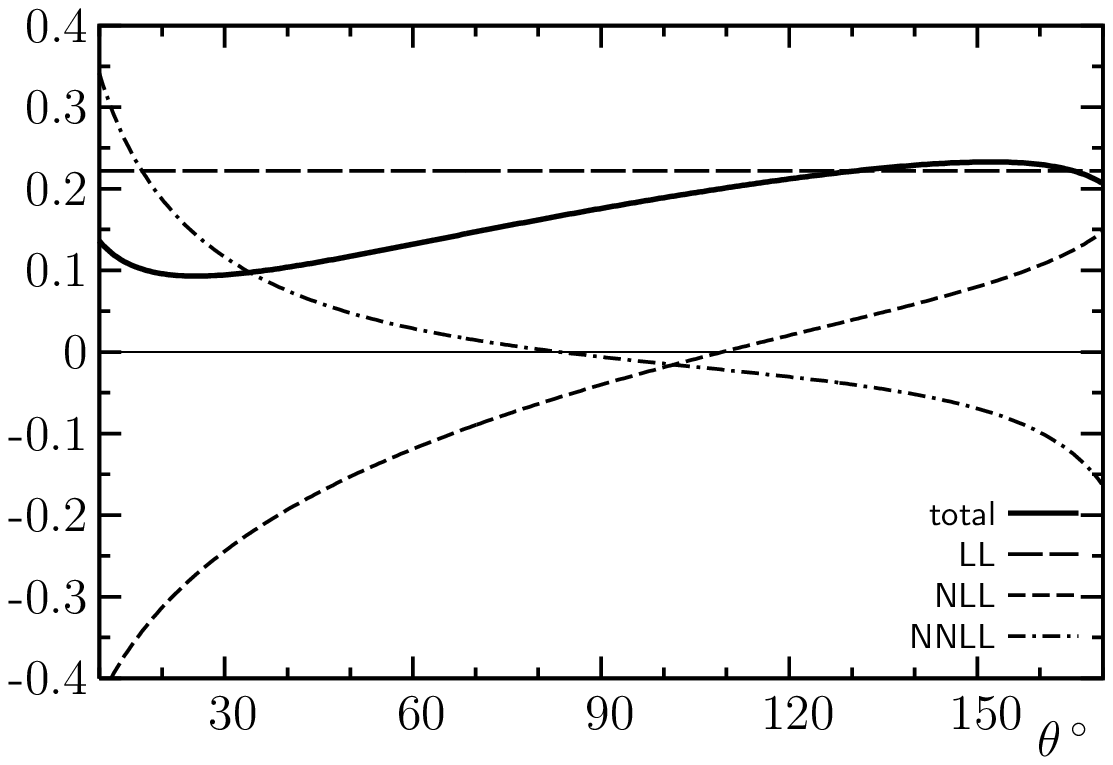,height=4.5cm}&
\hspace*{25pt}\epsfig{figure=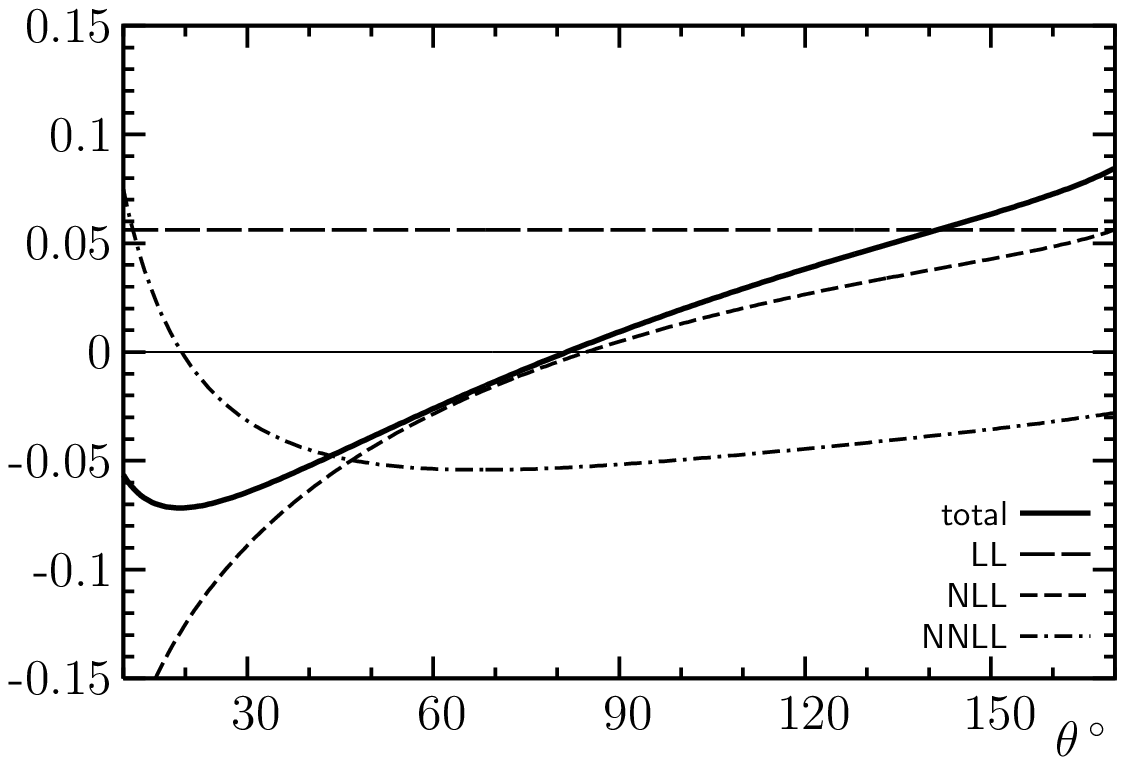,height=4.5cm}\\
\hspace*{-2pt}$(a)$ & \hspace*{47pt}$(b)$
\end{tabular}
\caption{\label{fig6} \small
  The same as Fig.~\ref{fig5} but for $\sqrt{s}=3$~TeV.}
\end{center}
\end{figure}

We adopt the following input values \cite{PDG} $M=M_W=80.41$~GeV,
$M_H=117$~GeV, $m_t=172.7$~GeV for the masses and
$\sin^2\theta_W=0.231$, $\al_{ew}=3.38\cdot 10^{-2}$ for the
$\overline{\rm MS}$ coupling constants renormalized at the scale of
the gauge boson mass. Note that the coupling constants
in the Born cross section of the longitudinal gauge boson production
are renormalized at the scale $\sqrt{s}$.
\vspace*{-3mm}
\paragraph{\bf\em  Transverse polarization.}
We obtain the following one and two-loop NNLL corrections to the
cross section
\begin{eqnarray}
\delta^{(1)}_{T}&=&-4.73\,{\cal L}^2(s) +\left[\left(-6.15 +
4.00\, {\xm \over \xp} \right)\ln(\xm) + 2.15\, \ln(\xp) + 4.43\right]
{\cal L}(s)
\nn\\
&&+\left(-{4.70 \over \xp}+{2.35+1.95\,\xm \over \xm^2+\xp^2} \right)
\ln^2(\xm)+{4.00 \over \xp}\ln(\xm)\ln(\xp)
+{3.00\,\xm \over \xm^2+\xp^2}\ln^2(\xp)
\nn\\
&&+\left(0.54+{4.95-9.65\,\xm\over \xm^2+\xp^2}\right)
\ln(\xm)-0.54\ln(\xp)-{2.35\,\xp\over \xm^2+\xp^2}-0.82\,,
\label{1lrest}
\\
\delta^{(2)}_{T}&=&11.17\,{\cal L}^4(s)
+\left[\left(29.08-{18.91\, \xm \over \xp}\right)\ln(\xm)
-10.17\,\ln(\xp)-14.62\right]{\cal L}^3(s) \nn\\&&
+\Bigg[\left(18.92+{22.21 -12.61\,\xm\over \xp}+{4.00\, \xm^2 \over
\xp^2} -{ 11.11+9.22\,\xm\over \xm^2+\xp^2}\right)\ln^2(\xm)
\nn\\
&&-\left(9.24+{18.91 \over \xp}
+{3.39\,\xm \over\xp}\right)\ln(\xm)\ln(\xp)
+\left(2.32-{14.18\,\xm \over \xm^2+\xp^2}\right)\ln^2(\xp)
\nn\\
&&-\left(15.26-{11.40\,\xm \over \xp}+ {23.40-45.61\,\xm \over
\xm^2+\xp^2}\right)\ln(\xm) +3.87\,\ln(\xp)+{11.11\,\xp\over \xm^2+\xp^2}
\nn\\
&&
-3.60\Bigg]{\cal
L}^2(s)\,.\label{2lrest}
\end{eqnarray}
The one-loop result is well known \cite{Bee} and
is given here to show the structure of the logarithmic expansion,
see Figs.~\ref{fig3}$a$ and \ref{fig4}$a$.
In Figs.~\ref{fig5}$a$ and \ref{fig6}$a$ the
values of different logarithmic two-loop contributions  as well as their
sum are plotted as functions of the production angle
at the center of mass energy of $1$~TeV and $3$~TeV, respectively.
The two-loop subleading contributions  exceed the LL one in absolute value in
the small angle region. However, due to the partial cancellation between
the NLL and NNLL terms the total NNLL approximation is  close to
the LL one. It has a fairly flat angular dependence
and amount to about 5\% for $\sqrt{s}= 1$~TeV
and 15\% for $\sqrt{s}= 3$~TeV.
\vspace*{-3mm}

\paragraph{\bf\em  Longitudinal polarization.}
For the left-handed initial state fermions the one and two-loop NNLL
corrections to the cross section read
\begin{eqnarray}
\delta^{(1)}_{L}&=&-2.38\, {\cal L}^2(s)+
\left[-6.91\,\ln(\xm)+0.75\,\ln(\xp)-3.48\right]{\cal L}(s)
\nn\\
&&-{2.19 \over \xp} \ln^2(\xm)+{0.65 \over \xm} \ln^2(\xp)
+0.19\,\left(\ln(\xm)-\ln(\xp)\right)+36.85\,,
\label{1lresl}
\\
\delta^{(2)}_{L}&=&2.82\,{\cal L}^4(s)
+\left[16.41\,\ln(\xm)-1.79\,\ln(\xp)+11.87\right]{\cal L}^3(s)
\nn\\
&&+\Bigg[\left(18.38+{5.20 \over \xp}\right) \ln^2(\xm)+
\left(0.28-{1.55 \over \xm}\right)\ln^2(\xp)
\nn\\
&&+3.11\,\ln(\xm)\ln(\xp)+49.(10.)\,\ln(\xm)-18.(4.)\,\ln(\xp)-128.(20.)\Bigg]
{\cal  L}^2(s)\,. \label{2lresl}
\end{eqnarray}
In the
two-loop NNLL contribution the error bars indicate the uncertainty
due to our approximation of the Yukawa enhanced contribution. The
structure of the logarithmic corrections differs from  the case
of the transverse polarization as one can see on
Figs.~\ref{fig3}$b$,  \ref{fig4}$b$ and \ref{fig5}$b$.
The sum of the  two-loop logarithmic terms is strongly angular dependent and
varies between -3\% and 2\% for $\sqrt{s}= 1$~TeV and
between -7\% and 8\% for $\sqrt{s}= 3$~TeV.

For the right-handed initial state fermions the Born cross section is
suppressed by the factor $4\sin^4\theta_W\approx 0.2$ in comparison to
the left-handed case. Moreover the two-loop logarithmic corrections
turned out to be about $3\cdot 10^{-3}$ for all the scattering angles.
Thus they are of no phenomenological importance and are not presented
here.

\section{Summary}
\label{sum} In the present paper we employed the evolution equation
approach to analyze the high energy asymptotic behavior of the gauge
boson production through the annihilation of fermion-antifermion
pair in the spontaneously broken $SU(2)$ gauge model.  The result
has been used to compute the two-loop NNLL electroweak corrections
to the differential cross section of $W$-pair production in $e^+e^-$
annihilation.  The corrections are comparable and even exceed the LL
terms depending on the production angle. The structure of the
corrections is different for the transverse and longitudinal boson
production.  In the first case we observe the cancellation between
the huge NLL and NNLL contributions so that the sum is dominated by
the LL term and amounts of about 5\% at $\sqrt{s}\sim 1$~TeV and
15\% at $\sqrt{s}\sim 3$~TeV.  For the longitudinal bosons the
corrections exhibit significant cancellation between the LL, NLL and
NNLL terms so that the sum does not exceed 2\% in absolute value for
$\sqrt{s}\sim 1$~TeV. The cancellation becomes less pronounced at
higher energy. The uncertainty of the theoretical prediction for the
on-shell $W$-pair production at ILC is now determined by the unknown
two-loop linear logarithmic terms. For the fermion pair production
such terms are know to contribute about 1--2\% of the cross section
\cite{JKPS}.  This value can be used as a rough  estimate of the
accuracy of our approximation.  We should emphasize that our
approximation breaks down at small production angles where the {\it
  Regge} logarithms $\ln(-t/s)$ become large.

\vspace{5mm}
\noindent
{\bf Acknowledgments}\\[3mm]
We thank S. Dittmaier and S. Uccirati for usefull discussion.
This work is supported by the DFG Sonder\-forschungsbereich/Transregio~9
``Computergest\"utzte Theoretische Teilchenphysik'' SFB/TR 9
and by BMBF grant 05HT4VKA/3. The work of F.M. has been supported by
the Graduiertenkolleg  "Hochenergiephysik und Teilchenastrophysik".

\end{document}